\documentclass[aps,prl,reprint,twocolumn,superscriptaddress,showpacs]{revtex4-1}
\usepackage{amsmath}
\usepackage{amssymb}
\usepackage{bbold}
\usepackage{bm}
\usepackage{braket}
\usepackage{graphicx}
\usepackage{hyperref}
\newcommand{\sgn}{\operatorname{sgn}}

\begin{document}
    \title{Valley-based Cooper Pair Splitting \\ via Topologically Confined 
        Channels in Bilayer Graphene}

    \author{Alexander Schroer}
    \affiliation{Institut f\"ur Mathematische Physik, Technische Universit\"at
        Braunschweig, D-38106 Braunschweig, Germany}
    \author{Peter G. Silvestrov}
    \affiliation{Institut f\"ur Mathematische Physik, Technische Universit\"at
        Braunschweig, D-38106 Braunschweig, Germany}
    \author{Patrik Recher}
    \affiliation{Institut f\"ur Mathematische Physik, Technische Universit\"at
        Braunschweig, D-38106 Braunschweig, Germany}
    \affiliation{Laboratory for Emerging Nanometrology Braunschweig, D-38106 
        Braunschweig, Germany}

    \begin{abstract}
        Bilayer graphene hosts valley-chiral one dimensional modes at domain
        walls between regions of different interlayer potential or stacking
        order. When such a channel is brought into proximity to a
        superconductor, the two electrons of a Cooper pair which tunnel into it
        move in opposite directions because they belong to different valleys
        related by the time-reversal symmetry. This is a kinetic variant of
        Cooper pair splitting, which requires neither Coulomb repulsion nor
        energy filtering but is enforced by the robustness of the valley isospin
        in the absence of atomic-scale defects. We derive an effective model for
        the guided modes in proximity to an $s$-wave superconductor, calculate
        the conductance carried by split and spin-entangled electron pairs, and
        interpret it as a result of \emph{local} Andreev reflection processes,
        whereas crossed Andreev reflection is absent.
    \end{abstract}

    % 72.80.Vp Electronic transport in graphene
    % 74.45.+c Proximity effects; Andreev reflection; SN and SNS junctions
    % 03.65.Ud Entanglement and quantum nonlocality (e.g. EPR paradox, Bell's
    %          inequalities, GHZ states, etc.) (for entanglement production and
    %          manipulation, see 03.67.Bg; for entanglement measures, witnesses
    %          etc., see 03.67.Mn; for entanglement in Bose-Einstein 
    %          condensates, see 03.75.Gg)
    \maketitle

    Creating mobile nonlocal spin-entangled electrons in a transport experiment
    with the help of superconductor--normal junctions has attracted a lot of
    attention in theory \cite{recher01,lesovik01,recher02,bena02,recher03,
    yeyati07,cayssol08,sato10} and experiment \cite{hofstetter09,herrmann10,
    schindele12,das12,tan15,deacon15} because the spin degree of freedom of the
    electron could serve as a solid-state qubit \cite{loss98}. In the existing
    experiments, the envisaged process where a Cooper pair is split over two
    normal leads is crossed Andreev reflection (CAR) \cite{torres99,falci01},
    which is enhanced by the repulsive electron-electron interaction on two
    quantum dots weakly coupled to the superconductor \cite{recher01} or by
    energy filtering \cite{lesovik01,sadovskyy14}. The basic mechanism of these
    entanglers is not very sensitive to the specific material used, i.e., the
    underlying band structure. It has been shown that characteristic features of
    new materials exhibiting Dirac-cones like graphene or topological insulators
    can be useful for splitting Cooper pairs \cite{cayssol08,chen11,
    reinthaler13,nilsson08,wang15}. In these proposals, the efficiency of the
    splitting process, in the absence of interactions, relies on non-protected
    resonance conditions or the split Cooper pair is not spin-entangled due to
    spin-helicity or spin-polarization of the leads.  Helical edge states of the
    quantum spin Hall regime have, however, been proposed to detect spin
    entanglement \cite{sato10,chen12}.

    \begin{figure}
        \includegraphics{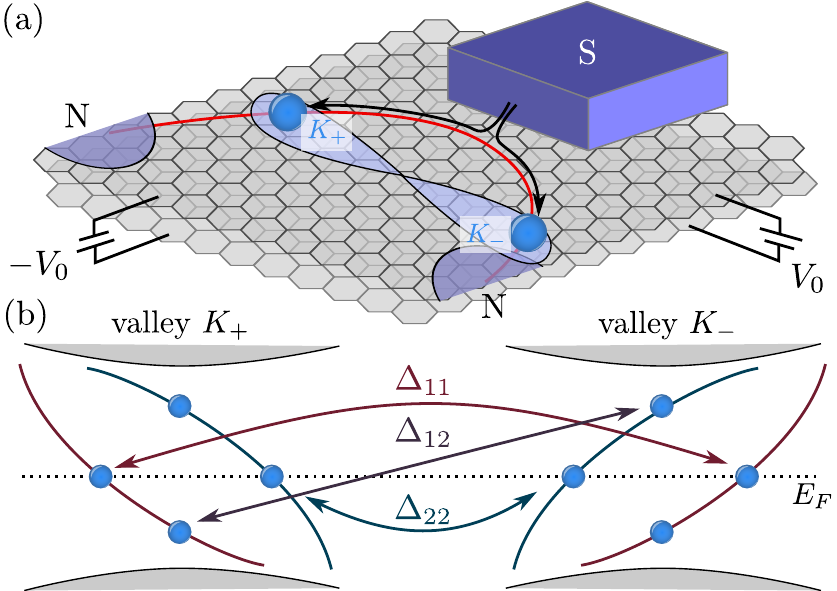}
        \caption{(a) At a domain wall (red) in BG between different interlayer 
            voltages $\pm V_0$ or different stacking order a topological
            valley-chiral channel forms. Cooper pairs tunneling into it from
            a nearby $s$-wave superconductor (S) are split because the two
            electrons belong to opposite valleys $K_\pm$ and thus have opposite
            velocities. They remain spin entangled and propagate to separate
            normal leads (N).  
            (b) In each valley two subgap modes along the domain wall emerge.
            Energy and momentum conservation along the NS interface single out
            four points in the subgap spectrum at which Cooper pairs are
            injected.}
        \label{fig-setup}
    \end{figure}

    Here, we propose to exploit the valley degree of freedom in bilayer graphene
    (BG), where valley-chiral, spin-degenerate one-dimensional (1D) channels
    are formed at domain walls. Such domain walls can be engineered by switching
    the sign of an interlayer voltage or by reversing the stacking order
    \cite{martin08,zhang13}. If brought into proximity to a superconductor, the
    pairs emitted into the channel are split, i.e., two electrons propagate in
    different directions but remain spin-entangled since, as required by
    time-reversal symmetry, the two electrons forming the Cooper pair in the
    superconductor are from different valleys \cite{beenakker08}. As long as the
    valley degree of freedom is robust, the splitting efficiency is unity,
    independent of resonance conditions. The device extends the upcoming 
    ``valleytronics'' in graphene \cite{rycerz07} to nonlocal
    Einstein-Podolsky-Rosen pairs.  A 1D channel defined by opposite stacking
    order has recently been created experimentally in BG \cite{ju15} with mean
    free paths over several 100~nm, demonstrating weak intervalley scattering.
    In this scenario, the normal reflection of an incoming hole (or electron)
    and CAR are absent.  In the limit of a weak proximity effect, where the
    normal transmission through the proximity region of the channel is almost
    perfect, the spin-entangled pair emission with electrons moving to opposite
    normal terminals (Fig.~\ref{fig-setup}) is equivalent to local Andreev
    reflection (LAR) processes, opposite to the normal reflection-dominated
    case, where CAR produces entangled pairs.

    We analyze the setup of Fig.~\ref{fig-setup} in two steps: first, we
    investigate the influence of the superconductor on the 1D channel by solving
    a Bogoliubov-de Gennes (BdG) equation, and derive an effective 1D model to
    describe the proximity effect in the channel. Second, we calculate the
    subgap conductance when applying a bias voltage between the superconductor
    and the channel using a rate equation approach. We interpret the subgap
    transport in a scattering matrix picture and show that the outgoing
    scattered state is a two-particle spin-entangled state on top of a filled
    normal-state Fermi sea with a chemical potential lowered by the bias
    voltage.  To leading order its weight is given by the LAR amplitude.

    \emph{Model.}---%
        We consider a BG sheet with Bernal $AB$ stacking in the presence of an
        interlayer voltage $V(\bm r)$ \cite{mccann06}. We model the
        superconductor region as BG in which the bands are shifted by a scalar
        potential $U(\bm{r})$ due to doping and which has an induced $s$-wave
        pairing amplitude $\Delta(\bm{r})$. We employ the low-energy
        approximation for BG, valid at energies and (inter)layer voltages
        smaller than the interlayer hopping $\gamma_1\simeq0.3$~eV. Without the
        superconductor, the valley index $\chi_v=\pm 1$ distinguishing the two
        $K$-points $K_\pm=\pm(4\pi/3a,0)$ and the electron spin
        $s=\pm1\equiv\uparrow,\downarrow$ are good quantum numbers and we write
        the Bogoliubov-de Gennes equation as
        $H_\text{BdG}^{\chi_v,s}\Phi_{\chi_v,s}({\bm r})=\varepsilon
        \Phi_{\chi_v,s}({\bm r})$
        \begin{align}
            H_\text{BdG}^{\chi_v,s}=&\left[
                \alpha\hbar^2\left(2\chi_v\partial_x\partial_y\sigma_y
                    +(\partial_x^2-\partial_y^2)\sigma_x\right)\right. \notag\\
                &\left.+U({\bm r})+V({\bm r})\sigma_z\right]\tau_z
                    +\Delta({\bm r})\tau_x,
                \label{eq-hbdg}
        \end{align}
        where $\alpha=v_F^2/\gamma_1$.  The Pauli matrices $\sigma_i$ act in the
        pseudospin $(A_1,B_2)$ space and $\tau_i$ in electron-hole space and we
        set the Fermi energy $E_F=0$.  The 4-component spinor is
        $\Phi_{\chi_v,s}({\bf r})=({\bm u}_{\chi_v,s}({\bm r}), {\bm
        v}_{\chi_v,s}({\bm r}))^{\rm T}$ where we have introduced the electron
        ${\bm u}_{\chi_v,s}({\bm r})=(u_{A_1,\chi_v,s}({\bm r}),
        u_{B_2,\chi_v,s}({\bm r}))$ and hole components ($u\rightarrow v$) on
        the two sublattices.  Excitations with energy $\varepsilon$ are then
        expanded as $\gamma_{\chi_v,s}(\varepsilon)= \int
        d^{2}r\,\Phi_{\chi_v,s}^{*}({\bf r})\cdot\Psi_{\chi_v,s}({\bf r})$ with
        the vector of field-operators $\Psi_{\chi_v,s}({\bf r})=
        (\psi_{A_1,\chi_v,s}({\bm r}),\psi_{B_2,\chi_v,s}({\bm r}),
        s\,\psi_{A_1,-\chi_v,-s}^{\dagger}({\bm r}),
        s\,\psi_{B_2,-\chi_v,-s}^{\dagger}({\bm r}))^{\rm T}$. 

        In the absence of the superconductor ($\Delta({\bm r})=U({\bm r})=0$)
        and assuming the modes to propagate along the $y$-direction along a
        domain wall at $x=0$, i.e., $V({\bm r})=-V_0\sgn(x)$ with $V_0>0$, the
        topologically confined modes can be found analytically \cite{martin08}.
        The electron and hole sectors in Eq.~\eqref{eq-hbdg} decouple. In the
        electron sector, the solutions in each half space have the form
        $\Phi_{\chi_v,s}({\bf r})= ({\bm u}^{0}_{\chi_v,s}({\bm r}),0)^{\rm T}$
        where ${\bm u}^{0}_{\chi_v,s}({\bm r})={\bm u}_{\chi_v,s}^{0}
        e^{\frac{i}{\hbar}(p_x x+p_y y)}$ with 
        \begin{equation}
             {\bm u}_{\chi_v,s}^{0}=\begin{pmatrix}
                -\varepsilon-V \\ \alpha^2(p_x+i \chi_v p_y)^2 
            \end{pmatrix}. 
            \label{eq-normal-psi0}
        \end{equation}
        For any fixed energy $\varepsilon$ and momentum $p_y$ there are four
        allowed values $p_x=\pm\sqrt{\pm i\sqrt{V_0^2-\varepsilon^2}/
        \alpha-p_y^2}$ which become complex when $|\varepsilon|<V_0$, i.e.
        there are no propagating modes in the bulk at energies below $V_0$.
        Matching the wavefunctions decaying away from the domain wall and their
        derivatives, one obtains the two electronic subgap solutions $n=1,2$ in
        each valley, $\Phi^{0,n}_{\chi_v,s,p_y}(x)$, with the dispersion
        relation
        \begin{equation}
            \varepsilon^{0,1/2}_{\chi_v,s,p_y}
            =\pm\frac{\sqrt{2}V_0-\alpha p_y^2}{2}
            -\frac{\chi_vp_y}{2}\sqrt{2\sqrt{2}\alpha V_0+\alpha^2 p_y^2},
            \label{eq-normal-e}
        \end{equation}
        with velocities opposite in the two valleys \cite{martin08}. The
        solutions for the hole sector
        $\Phi^{0,3/4}_{\chi_v,s,p_y}(x)=
        (0,{\bm v}^{0}_{\chi_v,s}({\bm r}))^{\rm T}$ where 
        ${\bm v}^{0}_{\chi_v,s}({\bm r})$ at energy
        $\varepsilon^{0,3/4}_{\chi_v,s,p_y}$ are obtained from 
        Eqs.~\eqref{eq-normal-psi0} and~\eqref{eq-normal-e} by setting
        $\varepsilon\rightarrow -\varepsilon$.

        The relevant momenta $p_y$ are close to the $K$ points: taking
        $\varepsilon\sim0$, we obtain from Eq.~\eqref{eq-normal-e} the momentum
        scale $p_y\sim\sqrt{V_0/\alpha}$, on which the K-points are located at 
        $\frac{4\pi\hbar}{3a}/\sqrt{V_0/\alpha}\sim10^2$ for
        $V\sim\Delta\sim\text{meV}$. The guided modes decay into the bulk on a
        length scale of $\sqrt{\hbar^2\alpha/V_0}$, which then is on the order
        of several $10$~nm. This sets the scale of the separation between the
        guided mode and a superconductor required to obtain a proximity effect. 

\begin{figure}
            \includegraphics{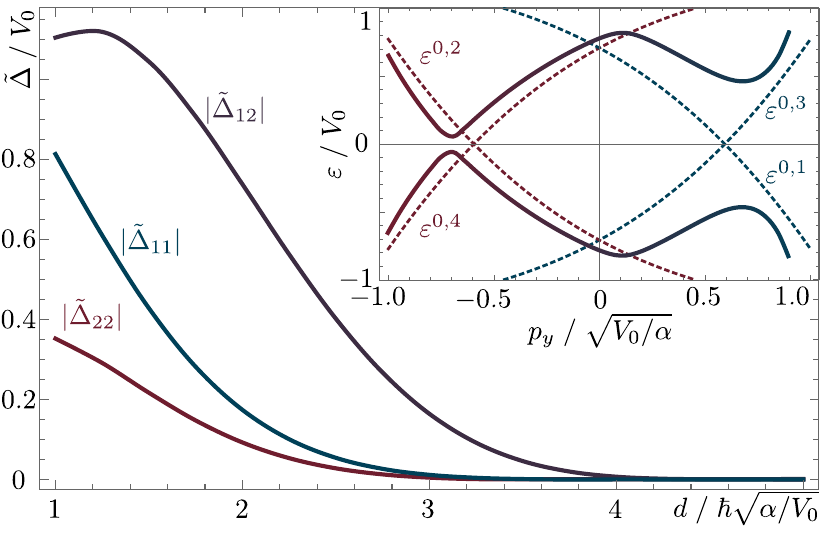}
            \caption{Induced intraband superconductivity $\tilde\Delta_{11}$
                ($\tilde\Delta_{22}$) in the 1D channel at the respective Fermi
                point $p_y=\pm2^{-\frac{3}{4}}\sqrt{V_0/\alpha}$ and induced
                interband superconductivity $\tilde\Delta_{12}$ at the band
                crossing $p_y=0$. For illustrative purposes we choose the bulk
                superconducting gap $\Delta=V_0$ and the doping $U_S=10V_0$ 
                ($\Delta<V_0$ is equally feasible). The amplitudes decay
                exponentially with the separation $d$ between the superconductor
                and the channel because the $V>0$ region acts as a tunnel
                barrier. Inset: In the normal state dispersion (dashed) two
                different-sized gaps open at the Fermi energy because
                $\tilde\Delta_{11}\neq\tilde\Delta_{22}$, shown for
                $d=1.5\sqrt{\hbar^2\alpha/V_0}$. Additionally,
                $\tilde\Delta_{12}$ opens a gap at the electron-hole crossing at
                $p_y=0$. This point contributes significantly to Cooper pair
                transport because compared to the Fermi points the normal
                density of states is higher and because the energy is larger
                such that the bound states extend further into the bulk,
                increasing the coupling to the superconductor.}
            \label{fig-pert}
        \end{figure}

    \emph{Perturbation theory for superconducting pairing.}---%
        Assuming a superconductor/normal interface with translational invariance
        along the $y$-direction, there are three distinct areas: in the
        superconductor area, $x<-d$, the pairing amplitude $\Delta({\bm r})=
        \Delta$ is finite and $U({\bm r})=-U_S$ is negative. The area
        $-d<x<0$ is in the normal state as before, $\Delta=U_S=0$, but the
        interlayer voltage is finite, $V=V_0>0$.  This region is a tunnel
        barrier between the superconductor and the domain wall at the interface
        to the third region $x>0$, where $\Delta=U_S=0$ and $V=-V_0$. In this
        situation guided modes exist at $|\varepsilon|<\min(V_0,\Delta)$ because
        states above $V_0$ can propagate in the normal regions and states above
        $\Delta$ can propagate in the superconductor.  Because of the tunnel
        barrier the guided modes are only weakly affected by the superconductor
        and we can apply standard quasidegenerate perturbation theory
        \cite{winkler03,suppl}, for which the unperturbed Hamiltonian $H_0$ is
        obtained from $H_\text{BdG}$ by setting $\Delta=U=0$ everywhere and so
        the perturbation $H_1$, which adds the missing parts, is finite only at
        $x<-d$. As a result of the perturbation the electron and hole states of
        the channel acquire a finite overlap $\tilde\Delta_{nn'}(p_y)$, where
        $n,n'$ label the subgap bands. This allows for particle number
        non-conserving processes, i.e., Cooper pair transport. To first order
        only the electron and hole states belonging to the same subgap band mix,
        $\tilde\Delta_{11}=\tilde\Delta_{22}$, $\tilde\Delta_{12}=0$. This
        agrees with the result one expects when introducing superconductivity
        phenomenologically by constructing the BdG equation directly from the
        guided modes with a uniform pairing $\tilde\Delta_{11}\tau_x$. The
        second order corrections, which take into account the modification of
        the wavefunctions due to the superconductor, however, reveal that the
        situation is different in the geometry we consider. The electron hole
        overlap differs in both bands, $\tilde\Delta_{11}\neq\tilde\Delta_{22}$,
        and band mixing is finite, $\tilde\Delta_{12}\neq0$
        (Fig.~\ref{fig-pert}) \cite{*[{See }] [{ for a similar effect in
            bilayer systems with different superconductivity in each layer.}]
            parhizgar14}. This is confirmed by the full dispersion relation
        of $H_\text{BdG}$ we obtain by matching the 4-component spinor and its
        derivatives at both interfaces numerically (Fig.~\ref{fig-pert}, inset):
        two gaps of different size open at zero energy ($\tilde\Delta_{11}$ and
        $\tilde\Delta_{22}$) and two gaps open at zero momentum where electron
        and hole states from different subgap bands cross
        ($\tilde\Delta_{12}=\tilde\Delta_{21}^*$). This means that there is
        Cooper pair transport at zero energy as well as at the finite energies
        $\pm V_0/\sqrt{2}$.

\begin{figure}
            \includegraphics{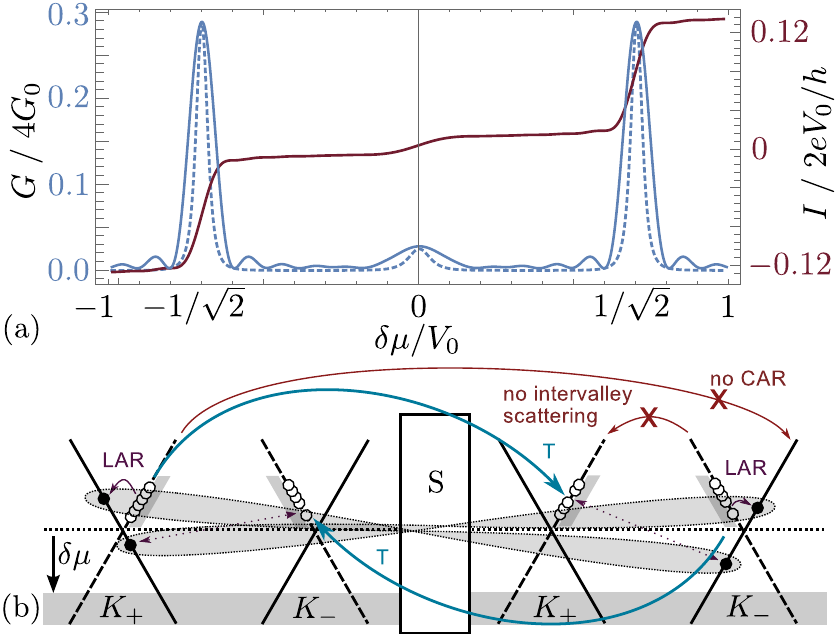}
            \caption{(a) Subgap conductance $G$ and Cooper pair current $I$ of a
                $w=25\hbar\sqrt{\alpha/V_0}$ long interface between the
                superconductor and the 1D channel at a distance
                $d=3\sqrt{\hbar^2\alpha/V_0}$ with $\Delta=U_S=V_0$. The peak
                structure reflects simultaneous energy and approximate momentum
                conservation. The oscillations are caused by the sharp boundary
                of the superconductor region and vanish if an exponential cutoff
                is used instead (dashed).  
                (b) Interpretation of Cooper pair splitting in terms of Andreev
                processes. Incoming holes (open circles) filled up to the bias
                $\delta\mu$ are either transmitted (T) or locally
                Andreev-reflected (LAR).  Ordinary reflection and crossed
                Andreev reflection (CAR) are zero by the valley chirality. A LAR
                process creates an outgoing electron (filled circle) on the same
                side and no outgoing hole on the opposite side, which
                corresponds to an electron of opposite spin, momentum and energy
                (dashed arrow). These two electrons are spin entangled (text).}
            \label{fig-conductance}
        \end{figure}

    \emph{Cooper pair transport.}---%
        We use Fermi's golden rule to calculate the Cooper pair current 
        $I=2e\sum_{fi}(W^+_{fi}-W^-_{fi})\rho_i$, where 
        $W^\pm_{fi}=\frac{2\pi}{\hbar}|\Braket{f_\pm|H_T|i}|^2
        \delta(\varepsilon_f-\varepsilon_i)$ is the transition rate from an
        initial state $i$ with probability $\rho_i$ at energy $\varepsilon_i$ to
        the final state $f_\pm$ with 2 more (less) electrons at energy
        $\varepsilon_f$. The tunnel Hamiltonian $H_T$ comprises the particle
        number non-conserving terms of the second-quantized perturbative model
        with electron operators $c_{\chi_v,s}^n(k)$ and hole operators
        $h_{\chi_v,s}^n(k)\equiv sc^{n\dagger}_{-\chi_v,-s}(-k)$, where $k\equiv
        p_y$,
        \begin{equation}
            H_T=\sum_{\chi_vnn'ks}\tilde\Delta_{nn'}(k)s
                c_{\chi_v,s}^n(k)c_{-\chi_v,-s}^{n'}(-k)+\text{H.c.}
            \label{eq-pairtunnel}
        \end{equation}
        Because the superconductor interface has a finite width $w$, we restrict
        the pairing amplitude in real space $\tilde\Delta(x,x')$ to 
        $x,x'\in[-w/2,w/2]$. In momentum space (suppressing all indices) this
        amounts to
        $\sum_k\tilde\Delta(k)c_kc_{-k}\longrightarrow\sum_{kk'} \Delta_{kk'}
        c_kc_{k'}$ with $\tilde\Delta(k)$ from the microscopic calculation and
        \begin{align*}
            \Delta_{kk'}&=
                \tilde\Delta\Big(\frac{k-k'}{2}\Big)
                \frac{L}{2\pi}\int dl
                \frac{\sin\Big[(l-k)\frac{w}{2}\Big]}{(l-k)\frac{L}{2}}
                \frac{\sin\Big[(l+k')\frac{w}{2}\Big]}{(l+k')\frac{L}{2}},
        \end{align*}
        where $L$ is the total length of the system, which does not enter the
        final results, and we have exploited that the integrand is peaked around
        $k\approx l \approx -k'$. With this, the rates for removing
        (adding) a Cooper pair, $\Ket{i}\rightarrow\Ket{f}=
        c_{\chi_v,s}^{n(\dagger)}(k)c_{-\chi_v,-s}^{n'(\dagger)}(k')\Ket{i}$,
        become
        $W^\mp_{fi}=4\pi|\Delta^{nn'}_{kk'}|^2
        \braket{\hat n_{\chi_v,s}^{n,e/h}(k)}_i
        \braket{\hat n_{-\chi_v,-s}^{n',e/h}(k')}_i
        \delta(\varepsilon_{\chi_v}^n(k)+\varepsilon_{-\chi_v}^{n'}(k'))$, where
        at low temperatures the occupation probability
        $\braket{\hat n_{\chi_v,s}^{n,e}(k)}_i=
        1-\braket{\hat n_{\chi_v,s}^{n,h}(k)}_i
        \approx\Theta(-\delta\mu-\varepsilon_{\chi_v}^n(k))$ with $\delta\mu$
        the voltage applied between the superconductor and the channel.
        Rewriting the sum over momenta as energy integrals, the current becomes
        \begin{align}
            I&=\frac{32e}{\hbar}\frac{\pi L^2}{(2\pi)^2}\sum_{nn'}
                \int_{-\delta\mu}^{\delta\mu}\hspace{-.25cm} d\varepsilon
                \Big|\frac{\partial k_{K_+n}(\varepsilon)}
                    {\partial\varepsilon}\Big|
                \Big|\frac{\partial k_{K_-n'}(-\varepsilon)}
                    {\partial\varepsilon} \Big| \notag\\
                &\hspace{1cm}\times
                |\Delta_{nn'}(\varepsilon,-\varepsilon)|^2.
        \end{align}
        The combination of energy conservation and approximate momentum
        conservation implies that the pair tunneling probability
        $|\Delta_{nn'}(\varepsilon,-\varepsilon)|^2$ has a single peak as a
        function of $\varepsilon$ for each pair $n,n'$. Injection into the
        same subgap band, $n=n'$, happens near $\varepsilon_0^{nn}=0$, and
        into different subbands, $n\neq n'$, near $\varepsilon_0^{12}=
        -\varepsilon_0^{21}=V_0/\sqrt{2}$ (Fig.~\ref{fig-setup}b).
        Linearizing the dispersion \eqref{eq-normal-e} around these points
        \cite{suppl}, $\varepsilon=\varepsilon_0^{nn'}+\hbar
        v_0^{nn'}(k-k_0^{nn'})$, the tunnel amplitude becomes
        $\Delta_{nn'}=\tilde\Delta_{nn'}(k_0^{nn'})\sin[
        (\varepsilon-\varepsilon_0^{nn'})(w/\hbar v_0^{nn'})]
        /L(\varepsilon-\varepsilon_0^{nn'})$ and we obtain the conductance
        \begin{align}
            G\approx4G_0\sum_{nn'}
                T_{nn'}
                \Big[\delta_w(\delta\mu-\varepsilon_0^{nn'})
                    +\delta_w(\delta\mu+\varepsilon_0^{nn'})
            \Big],
            \label{eq-cond}
        \end{align}
        where $G_0=2e^2/h$ is the conductance quantum,
        $\delta_w(\varepsilon)=\hbar v_0^{nn'}\sin^2 \Big[\varepsilon w/(\hbar
        v_0^{nn'})\Big]/(\pi w\varepsilon^2)$ becomes the delta function for
        $w\rightarrow\infty$, and $T_{nn'}=2\pi w
        |\tilde\Delta(k_0^{nn'})|^2/(\hbar v_0^{nn'})$ is the effective
        tunneling strength. Note that the conductance grows with the length of
        the interface. This is in contrast to conventional Cooper pair
        splitters, which suffer from an exponential suppression in the spatial
        size. The reason is that here Cooper pairs are split kinematically only
        after having tunneled locally into the channel, a process which can
        happen simultaneously along the whole interface. The conductance
        contains a central zero-bias peak and two characteristic side peaks
        [Fig.~\ref{fig-conductance}(a)], which arise because of the special
        subgap band structure and which correspond to the injection points
        marked in Fig.~\ref{fig-setup}(b). The peak height is proportional to
        the induced superconducting pairings. A factor of 4 arises due to the
        spin and valley degeneracy and a factor of 2 due to pair transport.

    \emph{Local Andreev reflection and Cooper pair splitting.}---%
        In Eq.~\eqref{eq-pairtunnel} the singlet nature of the injected Cooper
        pairs is manifest. It is well established that Cooper pair splitting is
        closely related to CAR \cite{prada04,samuelsson05}. This applies if the
        dominant process is ordinary reflection. In our device, the 1D channel
        with a proximity-induced superconducting region is a NSN junction, in
        which only transmission through the S-region with amplitude
        $t(\varepsilon)$ and \emph{local} Andreev reflection (an incoming
        quasiparticle in valley $K_\pm$ is reflected as an outgoing antiparticle
        with opposite velocity in valley $K_\mp$) with amplitude
        $r(\varepsilon)$ are possible. We consider the general scattering
        problem with finite CAR and normal reflection in the Supplemental
        Material \cite{suppl}.  When a voltage bias $\delta\mu$ is applied
        between the superconductor and the channel to extract Cooper pairs, the
        incoming modes are filled with holes up to $E_F+\delta\mu$. Without the
        superconductor, all are transmitted and fill the outgoing modes up to
        $E_F+\delta\mu$, which is equivalent to a Fermi sea for electrons
        $\Ket{}_{\delta\mu}$ with the Fermi energy $E_F-\delta\mu$ 
        \cite{samuelsson03,samuelsson05}
        \begin{equation}
            \Ket{}_{\delta\mu}\equiv
            \prod_{\substack{0<\varepsilon<\delta\mu\\\alpha s}}
                h^{\alpha\dagger}_s(\varepsilon)\Ket{}
            \equiv\prod_{\substack{0<\varepsilon<\delta\mu\\\alpha s}}
                c^\alpha_s(-\varepsilon)\Ket{}.
            \label{eq-newvac}
        \end{equation}
        Here, $\Ket{}$ is the quasiparticle vacuum with respect to the Fermi
        level $E_F$ of the superconductor and
        $h^{L/R\dagger}_s(\varepsilon)\equiv s c^{L/R}_{-s}(-\varepsilon)$
        creates outgoing holes with spin $s$ at energy $E_F+\varepsilon$ in the
        left/right lead, which is the same as annihilating an electron with
        opposite spin $-s$ at energy $E_F-\varepsilon$. We drop the valley index
        which is fixed by the requirement that outgoing modes move away from the
        superconducting region and the band index for simplicity \cite{suppl}.
        Due to the proximity effect LAR becomes finite. The key observation is
        that when LAR occurs, no hole with spin $s$ at energy $E_F+\varepsilon$
        is transmitted to the other side. The outgoing mode is therefore
        occupied by a spin $-s$ electron at energy $E_F-\varepsilon$
        [Fig.~\ref{fig-conductance}(b)]. To see this, we use 
        Eq.~\eqref{eq-newvac} to write the outgoing state in terms of
        $\Ket{}_{\delta\mu}$ \cite{suppl},
        \begin{align}
            &\prod_{\substack{0<\varepsilon<\delta\mu\\\alpha s}}
                \Big(s\,t\,c^{\bar\alpha}_{-s}(-\varepsilon)
                    +r\,c^{\alpha\dagger}_s(\varepsilon)\Big)
                c^{\bar\alpha\dagger}_{-s}(-\varepsilon)
                \Ket{}_{\delta\mu} \notag \\
            =&\prod_{\substack{0<\varepsilon<\delta\mu\\\alpha}}
                \Big[t^2
                    +r^2
                        c^{\alpha\dagger}_{\uparrow}(\varepsilon)
                        c^{\bar\alpha\dagger}_{\downarrow}(-\varepsilon)
                        c^{\alpha\dagger}_{\downarrow}(\varepsilon)
                        c^{\bar\alpha\dagger}_{\uparrow}(-\varepsilon)
                    \notag\\
                    \hspace{1cm}&+rt\Big(
                        c^{\alpha\dagger}_{\downarrow}(\varepsilon)
                        c^{\bar\alpha\dagger}_{\uparrow}(-\varepsilon)-
                        c^{\alpha\dagger}_{\uparrow}(\varepsilon)
                        c^{\bar\alpha\dagger}_{\downarrow}(-\varepsilon)
                        \Big)
                    \Big]\Ket{}_{\delta\mu}.
        \end{align}
        If $r$ is small, it becomes 
        $[1+\sum_{\varepsilon\alpha}r(
        c^{\alpha\dagger}_{\downarrow}(\varepsilon)
        c^{\bar\alpha\dagger}_{\uparrow}(-\varepsilon)-
        c^{\alpha\dagger}_{\uparrow}(\varepsilon)
        c^{\bar\alpha\dagger}_{\downarrow}(-\varepsilon))+\mathcal{O}(r^2)]
        \Ket{}_{\delta\mu}$, where the desired nonlocal singlet state is
        explicit. This corresponds to a situation, where individual splitting
        events are well separated and it is meaningful to talk about pairs. In
        this regime of interest the perturbative result from the previous
        section holds. Only the emitted pairs contribute to the shot noise of
        the scattering state.  In the opposite limit of perfect LAR with
        $\mathcal{O}(t)\sim0$, $\mathcal{O}(r)\sim 1$, the outgoing state is a
        nonentangled product state.  LAR is most pronounced at energies
        $\varepsilon=0$ and $\varepsilon=\pm V_0/\sqrt{2}$
        [Fig.~\ref{fig-conductance}(a)] where the superconductor opens gaps
        ${\tilde \Delta}_{nn'}$ in the spectrum for the case of an infinitely
        long ($w\rightarrow\infty$) tunnel-junction (Fig.~\ref{fig-pert}). The
        LAR process becomes weak for all energies, when $w$ falls below the
        coherence lengths $\hbar v_0^{nn'}/{\tilde \Delta}_{nn'}$.

    \emph{Conclusion.}---%
        Our setup allows for highly efficient creation of nonlocal
        spin-entangled electrons without the need for repulsive
        interaction or energy filters.  We note that the topological channel can
        be created electrically in the bulk of the BG sample, completely
        avoiding sharp sample edges, the main source of intervalley scattering
        \cite{dassarma11}, which could reduce the splitting efficiency.
        Moreover, using an electrically tunable channel geometry ballistic
        beamsplitters could be created to prove the spin entanglement via noise
        \cite{burkard00}, so far an elusive goal.  The spin relaxation and
        decoherence in BG are expected to be weak due to the small spin-orbit
        coupling \cite{gelderen10,guinea10,konschuh12} and the sparsity of
        nuclear spins.

    \begin{acknowledgments}
        We thank A.~Baumgartner, P.~Samuelsson, C.~Sch\"onenberger, and A.~Levy
        Yeyati for helpful discussions and acknowledge support from the
        EU-FP7 Project SE2ND, No. 271554, the DFG, Grant No. RE 2978/1-1 and
        Research Training Group GrK1952/1 ``Metrology for Complex Nanosystems'',
        and the Braunschweig International Graduate School of Metrology B-IGSM.
    \end{acknowledgments}

    \begin{appendix}
    \section{Supplemental Material}

    \setcounter{equation}{0}
    \renewcommand{\theequation}{A\arabic{equation}}
    \subsection{Perturbation theory for the proximity effect}
        Here we give the details of the calculation of the proximity induced
        amplitudes $\tilde\Delta$. Because the superconductor is only weakly
        coupled through the tunnel barrier $-d<x<0$, we can derive an effective
        1D model via low-order quasi-degenerate perturbation theory. We split 
        $H_\text{BdG}^{\chi_v,s}=H_0+H_1$ into two parts, where $H_0$ is
        diagonal in the eigenbasis $\Phi^{0,n}_{\chi_v,s,p_y}(x)$, and
        $H_1=\Theta(-x-d)[(-U_S-V_0\sigma_z)\tau_z+\Delta\tau_x]$. Since $H_1$
        is diagonal in spin and valley, we suppress the indices $\chi_v,s$ in
        the following. To first order in $H_1$,
        \begin{equation}
            H^{(1)}_{nn'}(p_y)=\int dx
                \Phi^{0,n\dagger}_{p_y}(x)H_1(x)\Phi^{0,n'}_{p_y}(x)
        \end{equation}
        and to second order,
        \begin{align}
            H^{(2)}_{nn'}(p_y)&=\frac{1}{2}\sum_{p_x}
                    \int dx\Phi^{0,n\dagger}_{p_y}(x)H_1(x)
                        {\tilde \Phi}^{0}_{p_x,p_y}(x) 
                        \notag\\&\hspace{.5cm}\times
                        \Big[(\varepsilon^{0,n}_{p_y}
                            -\tilde\varepsilon^{0}_{p_x,p_y})^{-1}
                        +(\varepsilon^{0,n'}_{p_y}
                            -\tilde\varepsilon^0_{p_x,p_y})^{-1}\Big]
                        \notag\\&\hspace{.5cm}\times
                        \int dx{\tilde\Phi}^{0\dagger}_{p_x,p_y}(x)H_1(x)
                        \Phi^{0,n'}_{p_y}(x),
        \end{align}
        where ${\tilde\Phi}^{0}_{p_x,p_y}(x)$ are the unperturbed free states 
        above the gap $(|\varepsilon|>V_0)$ with real $p_x$ and $p_y$ at
        energy ${\tilde\varepsilon}^{0}_{p_x,p_y}=
        \sqrt{\alpha^2(p_x^2+p_y^2)^2+V_0^2}$. We impose the quantization 
        condition $p_x=2\pi n/L$ and normalize the extended wavefunctions 
        according to $\int_{-L/2}^{L/2}dx{\tilde\Phi}^{0\dagger}_{p_x,p_y}(x)
        {\tilde\Phi}^{0}_{p_x,p_y}(x)=1$. The quantization length $L$ and the
        highest momentum $p_x$ are increased until the second order matrix 
        elements converge. To study Cooper pair transport only the parts of
        $H^{(1)}$, $H^{(2)}$ are relevant which are proportional to $\tau_x$,
        i.e., they mix electron and hole states and therefore change the
        particle number. The relevant momenta $p_y$ are close to the crossing of
        the respective electron and hole band (see the discussion on approximate
        momentum conservation in the main text). This can involve one band,
        $\tilde\Delta_{11}=H^{(1)}_{1,3}(p_F)+H^{(2)}_{1,3}(p_F)$ and
        $\tilde\Delta_{22}=H^{(1)}_{2,4}(-p_F)+H^{(2)}_{2,4}(-p_F)$, or both,
        $\tilde\Delta_{12}=\tilde\Delta^*_{21}=H^{(1)}_{1,4}(0)+
        H^{(2)}_{1,4}(0)$, where $\pm p_F$ are the Fermi points of the
        unperturbed dispersion, Eq.~\eqref{eq-normal-e}.

    \setcounter{equation}{0}
    \renewcommand{\theequation}{B\arabic{equation}}
    \vspace{1cm}
    \subsection{Linearized subgap dispersion}
        The linearized subgap dispersion, Eq.~\eqref{eq-normal-e}, reads
        \begin{equation}
            E\approx-\frac{4}{3}2^{1/4}\sqrt{V_0\alpha}
                \Big(\chi_v p_y\mp2^{-3/4}\sqrt{V_0/\alpha}\Bigr)
            \label{eq-normal-elin1}
        \end{equation}
        around the Fermi points and
        \begin{equation}
            E\approx\pm\frac{V_0}{\sqrt{2}}-\chi_v 2^{-1/4}\sqrt{V_0\alpha}p_y
            \label{eq-normal-elin2}
        \end{equation}
        around zero momentum. The coefficients $\varepsilon_0^{nn'}$ and
        $v_0^{nn'}$ used in the transport calculation, e.g.,
        Eq.~\eqref{eq-cond}, can be read off immediately. 

    \setcounter{equation}{0}
    \renewcommand{\theequation}{C\arabic{equation}}
    \subsection{Local and crossed Andreev reflection}
        In the most general case the incoming holes in a NSN junction can be
        transmitted $(t_{hh})$, reflected $(r_{hh})$, or undergo local
        ($r_{eh})$ or crossed ($t_{eh}$) Andreev reflection. The outgoing state
        is
        \begin{equation}
            \prod_{\varepsilon s\alpha}\Big(
                t_{hh}h^{\bar\alpha\dagger}_s(\varepsilon)
                +t_{eh}c^{\bar\alpha\dagger}_s(\varepsilon)
                +r_{hh}h^{\alpha\dagger}_s(\varepsilon)
                +r_{eh}c^{\alpha\dagger}_s(\varepsilon)\Big)\Ket{}.
        \end{equation}
        Rewriting the hole operators $h^\dagger$ in terms of electron operators 
        $c$, and the Fermi sea $\Ket{}$ in terms of the lowered Fermi sea
        $\Ket{}_{\delta\mu}$ as explained in Eq.~\eqref{eq-newvac} in the main 
        text, we arrive at
        \begin{widetext}
        \begin{align}
            \prod_{\varepsilon s}\Big[
                &(t_{hh}^2+r_{hh}^2)
                +(r_{eh}^2-t_{eh}^2)
                    c^{R\dagger}_s(\varepsilon)
                    c^{R\dagger}_{-s}(-\varepsilon)
                    c^{L\dagger}_s(\varepsilon)
                    c^{L\dagger}_{-s}(-\varepsilon) \notag\\
                +&s(r_{eh}r_{hh}-t_{hh}t_{eh})
                    (c^{L\dagger}_s(\varepsilon)c^{L\dagger}_{-s}(-\varepsilon)
                    +c^{R\dagger}_s(\varepsilon)c^{R\dagger}_{-s}(-\varepsilon))
                    \notag\\
                +&s(t_{eh}r_{hh}-t_{hh}r_{eh})
                    (c^{R\dagger}_s(\varepsilon)c^{L\dagger}_{-s}(-\varepsilon)
                    +c^{L\dagger}_s(\varepsilon)c^{R\dagger}_{-s}(-\varepsilon))
            \Big]\Ket{}_{\delta\mu}.
        \end{align}
        \end{widetext}
        The first line contains the product state contributions, the second line
        local pairs, and the third line nonlocal pairs. In the 
        conventional reflection-dominated case, $r_{hh}\sim1$, realized in
        Y-junction Cooper pair splitters, the leading order contributions are
        \begin{align}
            \Big[1+\sum_{\varepsilon s}\Big(&r_{eh}s
                    (c^{L\dagger}_s(\varepsilon)c^{L\dagger}_{-s}(-\varepsilon)
                    +c^{R\dagger}_s(\varepsilon)c^{R\dagger}_{-s}(-\varepsilon))
                    \notag\\+&t_{eh}s
                    (c^{R\dagger}_s(\varepsilon)c^{L\dagger}_{-s}(-\varepsilon)
                    +c^{L\dagger}_s(\varepsilon)c^{R\dagger}_{-s}(-\varepsilon))
            \Big)\Big]\Ket{}_{\delta\mu},
        \end{align}
        i.e., LAR produces local pairs and CAR produces nonlocal pairs. In the
        transmission-dominated situation, $t_{hh}\sim1$, the situation is
        reversed: the leading order is
        \begin{align}
            \Big[1-\sum_{\varepsilon s}\Big(&t_{eh}s
                    (c^{L\dagger}_s(\varepsilon)c^{L\dagger}_{-s}(-\varepsilon)
                    +c^{R\dagger}_s(\varepsilon)c^{R\dagger}_{-s}(-\varepsilon))
                    \notag\\+&r_{eh}s
                    (c^{R\dagger}_s(\varepsilon)c^{L\dagger}_{-s}(-\varepsilon)
                    +c^{L\dagger}_s(\varepsilon)c^{R\dagger}_{-s}(-\varepsilon))
            \Big)\Big]\Ket{}_{\delta\mu},
        \end{align}
        so LAR produces nonlocal pairs and CAR produces local pairs.  In the
        situation discussed in the main text, both CAR and reflection are
        forbidden, ruling out local pairs to all orders, as long as the valley
        symmetry is obeyed.  Generally speaking it is undesirable to have
        simultaneously strong ordinary reflection and strong LAR or to have
        simultaneously strong transmission and strong CAR to build a Cooper pair
        splitter useful to create spin entanglement.

    \setcounter{equation}{0}
    \renewcommand{\theequation}{D\arabic{equation}}
    \subsection{Multiband Andreev reflection}
        The notation becomes more cumbersome, when both subgap bands are
        considered but the considerations are completely analogous. Without
        superconductivity the outgoing scattering state is
        \begin{equation}
            \prod_{\substack{0<\varepsilon<\delta\mu\\
                    \alpha sn}}
                h^{\alpha\dagger}_{ns}(\varepsilon)\Ket{}
            \equiv\prod_{\substack{0<\varepsilon<\delta\mu\\\alpha sn}}
                c^\alpha_{ns}(\varepsilon)\Ket{}
            \equiv\Ket{}_\mu,
        \end{equation}
        where $n\in\{1,2\}$ is the band index. In the presence of the 
        superconductor, the transmitted holes can change the subgap band from
        $n$ to $m$ with an amplitude $t_{mn}(\varepsilon)$. Like in the one-band
        case, whenever the energy of an incoming electron is such that the
        spectrum of the S region has a gap, the transmission amplitude
        $t_{nm}(\varepsilon)$ is exponentially suppressed with the length of the
        proximity region, and due to unitarity there is a finite amplitude
        $r_{nm}(\varepsilon)$  for the spin-$s$ hole to be Andreev reflected
        locally as a spin-$s$ electron at energy $E_F+\varepsilon$:
        \begin{widetext}
        \begin{align}
            \Ket{\text{out}}&=\prod_{\substack{0<\varepsilon<\delta\mu
                    \\\alpha ns}}
                \sum_{m}
                \Big(s\,t_{mn}\,
                        c^{\bar\alpha}_{m,-s}(-\varepsilon)
                    +r_{mn}\,
                        c^{\alpha\dagger}_{ms}(\varepsilon)\Big)\Ket{}
                    \notag\\
            &=\prod_{\substack{0<\varepsilon<\delta\mu\\\alpha ns}}
                \sum_{m}
                \Big(s\,t_{mn}\,
                        c^{\bar\alpha}_{m,-s}(-\varepsilon)
                    +r_{mn}\,
                        c^{\alpha\dagger}_{ms}(\varepsilon)\Big)
                \prod_{m'}c^{\bar\alpha\dagger}_{m',-s}(-\varepsilon)
                \Ket{}_{\delta\mu} \notag \\
            &=\prod_{\substack{0<\varepsilon<\delta\mu\\\alpha s}}
                \Big[(t_{12}t_{21}-t_{11}t_{22})
                    +(r_{11}r_{22}-r_{12}r_{21})
                        c^{\alpha\dagger}_{1s}(\varepsilon)
                        c^{\alpha\dagger}_{2s}(\varepsilon)
                        c^{\bar\alpha\dagger}_{1,-s}(-\varepsilon)
                        c^{\bar\alpha\dagger}_{2,-s}(-\varepsilon)
                    \notag\\&\hspace{1cm}
                    +\sum_{nm}(-1)^m(r_{n1}t_{\bar m 2}-r_{n2}t_{\bar m 1})
                        s
                        c^{\alpha\dagger}_{ns}(\varepsilon)
                        c^{\bar\alpha\dagger}_{m,-s}(-\varepsilon)
                    \Big]\Ket{}_{\delta\mu}.
            \end{align}
        \end{widetext}
        Writing out the spin part of the product explicitly, the nonlocal 
        singlet nature of the injected Cooper pairs becomes obvious:
        \begin{widetext}
        \begin{align}
            \Ket{\text{out}}&=\prod_{-\delta\mu<\varepsilon<\delta\mu}
                \Big[(t_{12}t_{21}-t_{11}t_{22})^2
                    \notag\\&\hspace{2cm}
                    +(t_{12}t_{21}-t_{11}t_{22})
                    \sum_{nm}(-1)^m(r_{n1}t_{\bar m 2}-r_{n2}t_{\bar m 1})
                        (c^{L\dagger}_{n\uparrow}(\varepsilon)
                        c^{R\dagger}_{m\downarrow}(-\varepsilon)
                        -c^{L\dagger}_{n\downarrow}(\varepsilon)
                        c^{R\dagger}_{m\uparrow}(-\varepsilon))
                    \notag\\&\hspace{2cm}
                    +\mathcal{O}(r^2)
                \Big]\Ket{}_{\delta\mu}.
            \label{eq-out}
        \end{align}
        \end{widetext}
        The higher order terms in $r$ contain multiple Cooper pairs and are not
        necessarily entangled, e.g., the $\mathcal{O}(r^4)$ contribution is a
        pure product state in which all states in the left/right lead at energy 
        $E_F\pm\varepsilon$ are occupied. 
    \end{appendix}

\end{document}